\documentclass[journal=jpcafh,manuscript=article,chaptertitle=true,etalmode=truncate,maxauthors=10]{achemso}

\usepackage{graphicx}
\usepackage{dcolumn}
\usepackage{bm}
\usepackage{braket} 
\usepackage{mathtools}
\usepackage[utf8]{inputenc}
\usepackage[T1]{fontenc}
\usepackage{mathptmx}
\usepackage{amsmath}
\usepackage{hyperref}
\usepackage{color}

\title{Understanding the Linewidth of the ESR Spectrum Detected by a Single NV Center in Diamond} 

\author{Benjamin Fortman}
\affiliation{Department of Chemistry, University of Southern California, Los Angeles CA 90089, USA}

\author{Susumu Takahashi}
\affiliation{Department of Chemistry, University of Southern California, Los Angeles CA 90089, USA}
\alsoaffiliation{Department of Physics \& Astronomy, University of Southern California, Los Angeles CA 90089, USA}
\email{susumu.takahashi@usc.edu}
\begin{document}

\begin{abstract}
Spectral analysis of electron spin resonance (ESR) is a powerful technique for various investigations including characterization of spin systems, measurements of spin concentration, and probing spin dynamics.
The nitrogen-vacancy (NV) center in diamond is a promising magnetic sensor enabling improvement of ESR sensitivity to the level of a single spin.
Therefore, understanding the nature of NV-detected ESR (NV-ESR) spectrum is critical for applications to nanoscale ESR.
Within this work we investigate the linewidth of NV-ESR from single substitutional nitrogen centers (called P1 centers).
NV-ESR is detected by a double electron-electron resonance (DEER) technique.
By studying the dependence of the DEER excitation bandwidth on NV-ESR linewidth, we find that the spectral resolution is improved significantly and eventually limited by inhomogeneous broadening of the detected P1 ESR.
Moreover, we show that the NV-ESR linewidth can be as narrow as 0.3 MHz.
\end{abstract}

\maketitle

\section{Introduction}
Electron spin resonance (ESR) spectroscopy is a powerful technique to investigate properties of magnetic systems and their local environments.~\cite{abragram1970EPRtransitionmetals, schweiger2001principles, poole1983electron, Weil2007Electron}
In particular, ESR spectral analysis, in which the position, intensity, and lineshape of the ESR spectrum is carefully analyzed to extract spin parameters including g-values, hyperfine and spin-spin couplings, zero-field splittings and rotational correlation times of systems, is widely and routinely used for characterizations and investigations in science and engineering fields.
Examples include identification of paramagnetic defect contents in semiconductors,~\cite{Feher59, Watkins98, Loubser1978} investigations of structures and conformational dynamics of biological molecules,~\cite{Hubbell2000, freed11, britt15} and characterizations of photochemical reactions. ~\cite{puluektov17, waseilewski03, britt00}

The nitrogen-vacancy (NV) center is of significant interest in quantum sensing due to its unique properties.~\cite{Gruber1997, Degen2008, Balasubramanian2008, Maze2008}
The NV center is an $S = 1$ spin system consisting of a single vacancy site located adjacent to a nitrogen atom in the diamond lattice.
The electronic structure of the NV center enables optical readout and initialization of the spin system through optically detected magnetic resonance (ODMR).~\cite{Gruber1997} In addition, the long-lived quantum coherence of the NV centers' spin states,~\cite{balasubramanian08, Takahashi2008} provides the NV center with high sensitivity to external magnetic fields, with fields as small as 100 fT being reported.~\cite{Wolf2015}
Its atomic size combined with high sensitivity to magnetic fields gives the NV center a sensing radius in the nanometer range, allowing ESR sensitivity to improve to the level of a single electron spin.~\cite{Grinolds2013}
Using a single NV center, nanoscale ESR detection of several types of spins in solid state and biological systems has been demonstrated.~\cite{Sushkov2014,Grinolds2013,Abeywardana2016,Shi2013,Shi2015,Mamin2012,DeLange2012,Rosenfeld2018, meriles12}
For applications of NV-detected ESR (denoted NV-ESR in this manuscript) spectral analysis, it is critical to understand the nature of NV-ESR lineshape and to establish a procedure to obtain a high resolution spectrum representing intrinsic properties of the sample.

Here we investigate the nature of an NV-ESR spectrum of single substitutional nitrogen defects in diamond (called P1 centers).
The NV-ESR spectrum is obtained using a double electron-electron resonance (DEER) pulse sequence, which utilizes pulses at two distinct microwave (MW) frequencies to coherently control the NV center and target spins.
By studying the spectral linewidth as a function of the DEER pulse length, we identify a significant contribution of the DEER excitation bandwidth to the observed NV-ESR linewidth at short pulse lengths.
At long pulse lengths, we observe that the ESR linewidth is limited by inhomogeneous broadening of the detected P1 ESR frequency ($T_2^*$-limit), representing intrinsic spin dynamics of P1 spins.
Moreover, by employing a long DEER pulse, we observed that the ESR linewidth is as narrow as 0.3 MHz and, with the improvement of the spectral resolution, we clearly resolve a small splitting (2 MHz) in P1 ESR that originates from the anisotropic hyperfine coupling and four different orientations of the P1 spins.

\section{Materials and Methods}
\subsection{Diamond sample}
A single crystal (2.0 $\times$ 2.0 $\times$ 0.3 mm$^3$) of (111)-cut high pressure high temperature type-Ib diamond (purchased from Sumitomo electric industries) was used in this study.

\subsection{115 GHz ESR spectroscopy}
The 230 GHz/115 GHz ESR system employs a high-power solid-state source consists of a 9-11 GHz synthesizer, pin switch, microwave amplifiers, and frequency multipliers. The output power of the source system is 100 mW at 230 GHz and 700 mW at 115 GHz.
The 230/115 GHz excitation is propagated using a quasioptical bridge and a corrugated waveguide and couples to a sample located at the center of a 12.1 T cryogenic-free superconducting magnet.
ESR signals are isolated from the excitation using induction mode operation.~\cite{Smith98}
For ESR detection, we employ a superheterodyne detection system in which 115 GHz is down-converted into the intermediate frequency (IF) of 3 GHz then down-converted again to in-phase and quadrature components of dc signals. Details of the system have been described elsewhere.~\cite{Cho2014,Cho2015}
In the present experiment, the magnetic field modulation strength was adjusted to maximize the intensity of ESR signals without distorting the lineshape (typical modulation amplitude of 0.01 mT with modulation frequency of 20 kHz).

\subsection{ODMR spectroscopy}
The ODMR system is based on a homebuilt confocal microscope system.
A 100-mW 532-nm laser (Crystalaser) is passed through an acousto-optic modulator (Isomet 1250C) before being directed through a low-pass filter (Omega) and into a single mode fiber (Thorlabs).
The output of the fiber is directed through a dichroic mirror and up through a microscope objective (Zeiss 100X) to the sample stage.
Fluorescence (FL) is detected by an avalanche photodiode (Excelitas) through a high-pass filter (Omega) and another single mode fiber.
The autocorrelation measurement is performed with a Hanbury Brown-Twiss interferometer.~\cite{Brown1956}
For ODMR, microwave (MW) excitation is directed from the sources (Stanford Research Systems SG386 and Rohde-Schwarz SML03) through a power combiner, and a high gain amplifier to the sample stage.
A 20 $\mu$m gold wire is placed on the surface of the diamond for MW excitation and coherent control of the NV centers.

\section{Results and Discussion}
\begin{figure}
    \centering
    \includegraphics[scale=1.5]{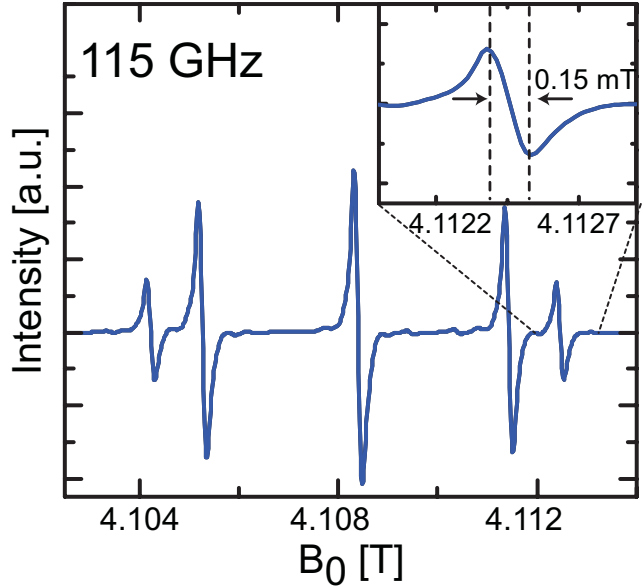}
    \caption{CW ESR spectrum of P1 centers taken at 115 GHz at room temperature. The inset graph shows the spectrum of the  $m_I=-1$ ESR signal. The ESR linewidth is $0.15 \pm 0.02$ mT. A field modulation of 0.01 mT at 20 kHz and a field sweep rate of $0.01$ mT/s were used.}
    \label{fig:HF_ESR}
\end{figure}
We first perform 115 GHz continuous wave ESR (CW ESR) spectroscopy to identify impurity contents within the diamond sample.
Figure \ref{fig:HF_ESR} shows CW ESR data of the diamond sample with application of an external magnetic field along the [111] direction.
We observe five pronounced ESR signals from P1 centers ($S = 1/2$, $I = 1$, ${A_{x,y}} = 82$ MHz and ${A_z} = 114$ MHz).~\cite{Loubser1978}
The intensity of 115 GHz wave excitation was reduced to avoid the saturation of the ESR signal and the intensity of magnetic field modulation was carefully adjusted to maximize the signal-to-noise ratio of the ESR signal without distortion.
The P1 spectrum consists of five ESR signals due to the four possible orientations of P1 and the anisotropic hyperfine coupling.
Namely, the signals at 4.104, 4.108 and 4.113 T correspond to the ESR of P1 centers oriented along the [111] direction while the signals at 4.105, 4.108 and 4.112 T are from the other three orientations, [$\Bar{1}11$], [$1\Bar{1}1$], and [$11\Bar{1}$].
As shown in the inset of Fig.~\ref{fig:HF_ESR}, the linewidth of the observed ESR is $0.15 \pm 0.02$ mT for the signal at 4.113 T.

\begin{figure}
\includegraphics[scale=1.2]{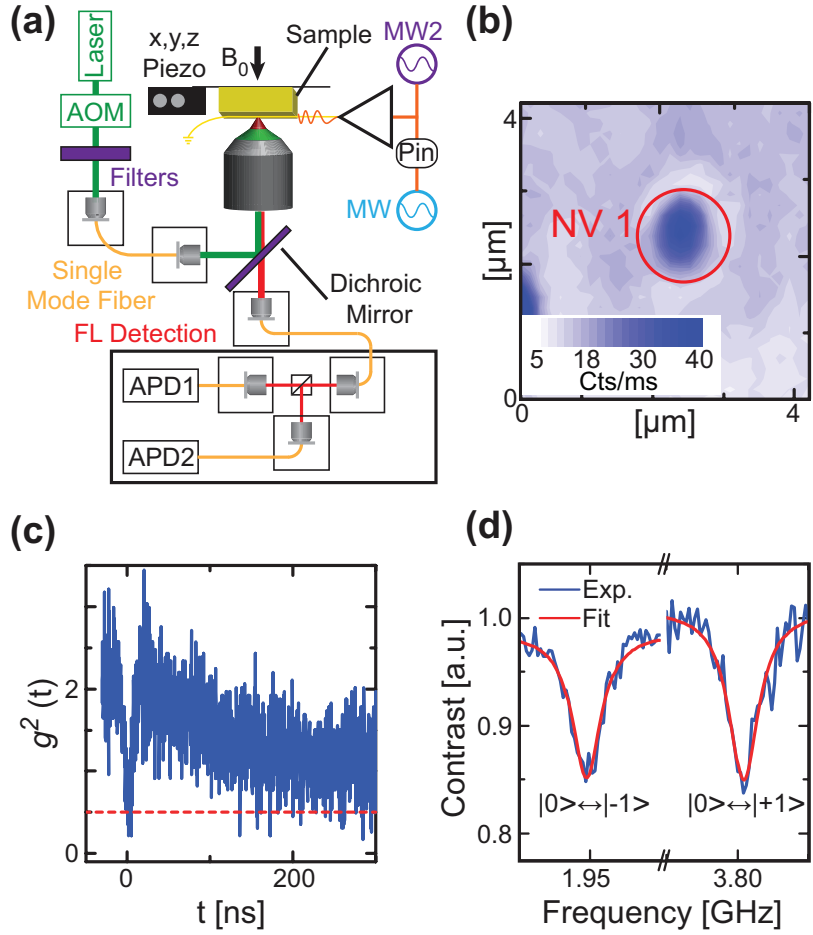}
\caption{ODMR experiment to identify a NV center.
(a) Diagram of the experimental setup.
(b) Spatial FL image with NV1 indicated in the solid red circle.
(c) Autocorrelation measurement of NV1.
The dotted red line drawn at 0.5 indicates the threshold for single quantum emitters.
(d) CW ODMR signals from the lower ($\ket{0} \leftrightarrow  \ket{-1}$) and upper ($\ket{0} \leftrightarrow  \ket{+1}$) transitions.
The signal is normalized by the FL intensity without MW excitation.}
\label{fig:Flimage}
\end{figure}
We next measure P1 ESR using a single NV center in diamond.
For NV-ESR measurements, we employ a homebuilt ODMR system as shown in Fig.~\ref{fig:Flimage}(a).
Figure~\ref{fig:Flimage}(b) shows a FL image of the diamond sample as well as an isolated FL peak for the present experiment (denoted as NV1).
An autocorrelation measurement, as shown in Fig.~\ref{fig:Flimage}(c), shows a dip in the signal at $t = 0$ that proves the FL emission is from a single quantum emitter.
CW ODMR measurements, where the FL intensity is monitored while sweeping the microwave frequency, are then performed on the isolated FL spot, as shown in Fig.~\ref{fig:Flimage}(d).
The observed ODMR signals correspond to the $m_S = -1$ and $m_S = 0$, and the $m_S = +1$ and $m_S = 0$ transitions of the NV center ($S = 1$, $g = 2.0028$ and $D = 2.87$ GHz). Therefore, we determine this spot to be a single NV center.
From the observed ODMR frequencies we determined the applied magnetic field to be 33.4 mT with a polar angle of $6.1 \pm 0.1$ degrees from the [111] axis.

\begin{figure}
\includegraphics[scale=1.5]{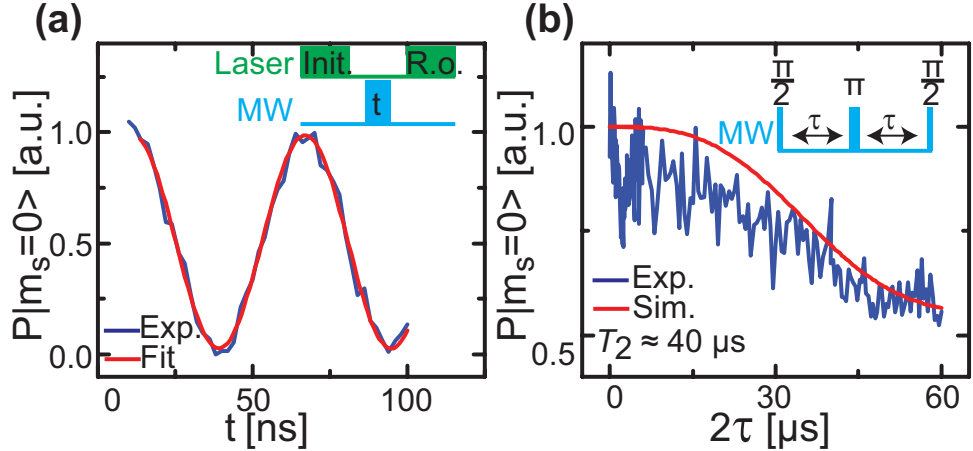}
\caption{Pulsed ODMR data collected from NV1 at 33.4 mT.
(a) Measurement of Rabi oscillations.
The Rabi oscillations show a $\pi$ pulse time of 40 ns.
The pulse sequence is shown in the inset.
The data is normalized to reflect the probability of the NV center being in the $m_S=0$ state (P$\ket{m_S=0}$).\cite{Abeywardana2016}
For all pulsed ODMR presented, a 5 $\mu$s laser pulse is used to initialize the spin state while a 300 ns laser pulse is used for readout.
Microwave pulses (shown as blue rectangles) are applied to drive the $\ket{0} \leftrightarrow \ket{-1}$ transition.
Each pulse sequence is repeated 10$^4$-10$^6$ times for an unweighted averaging of each data point.
(b) Spin echo measurement.
The spin echo data shows a spin decoherence time ($T_2$) of 40 $\mu$s for NV1.
Data is shown in agreement with  $I(t) = exp[-(t/T_2)^3]$.\cite{DeLange2010, Wang2013, Abeywardana2016}}
\label{fig:DataSE}
\end{figure}
\begin{figure}
\includegraphics[scale=1.5]{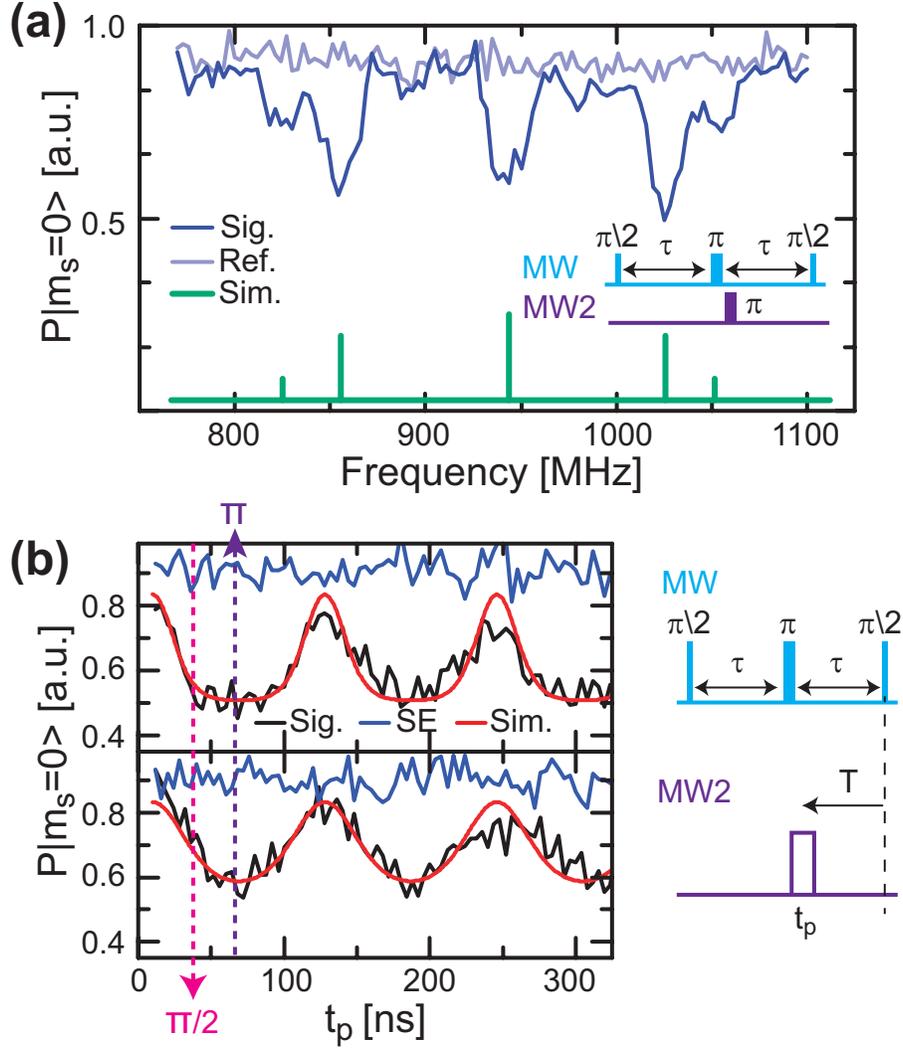}
\caption{NV-ESR of NV1 with $\tau = 5.7$ $\mu$s.
(a) NV-ESR spectrum obtained for NV1 with a 56-ns MW2 $\pi$-pulse.
SE intensity at the same $\tau$ is shown as a reference.
The DEER pulse length was chosen to maximize signal for the axial P1 orientation.
ESR frequencies calculated from the P1 spin Hamiltonian are shown in the stick spectrum.
(b) Rabi oscillations of P1 centers measured by NV-ESR.
The NV-ESR signal is plotted against $t_p$ (solid black).
The distance to the end of the sequence is indicated by $T$ and was 5.7 (2.0) $\mu$s for the upper (lower) data.
SE data (Solid blue) is shown as a reference.
The simulation using Eq.~\ref{eq:deerIntViktor} is shown in red.
The pulse sequence used for the NV-ESR Rabi experiment is shown to the right.}
\label{fig:DEER}
\end{figure}
Next, we perform the NV-ESR experiment.
We first conduct Rabi oscillation and spin echo (SE) measurements to determine pulse lengths and the spin coherence time ($T_2$) for NV-ESR.
As shown in the inset of Fig.~\ref{fig:DataSE}(a), the Rabi measurement is performed by first initializing the spin state into $\ket{0}$ with a long laser pulse before applying a variable length MW pulse.
The final spin state is then read out using a short laser pulse to induce FL from the NV center (see the inset of Fig.~\ref{fig:DataSE}(a)).
$T_2$ is measured using a Hahn spin echo sequence (see the inset of Fig.~\ref{fig:DataSE}(b)).
We determined the $T_2$ of NV1 to be 40 $\mu$s.
After the Rabi and SE experiments, we perform NV-ESR using a DEER technique, as shown in Fig.~\ref{fig:DEER}(a).
NV-ESR is performed by measuring the change in a coherent state of the NV center as a function of the frequency of the DEER pulse.
The coherent state change is induced by a shift of the magnetic dipole field of target spins due to the population inversion of target spins induced by the DEER pulse.
For this measurement, a $\tau$ of 5.7 $\mu$s was chosen to reduce decoherence of the NV center.
As shown in Fig.~\ref{fig:DEER}(a), the resulting spectrum exhibits five peaks, in agreement with P1 ESR.
In the measurement, the MW intensity is adjusted to ensure the DEER pulse length performs a $\pi$ rotation of the P1 center spins.
Figure~\ref{fig:DEER}(b) shows Rabi oscillations of P1 centers measured by NV-ESR with different $T$ values (see the sequence in Fig.~\ref{fig:DEER}(b)).
To explain the results, we consider the following NV-ESR model which describes the spin dynamics of an ensemble of two-level systems.~\cite{Stepanov2016}
Using this model, the intensity of NV-ESR is given by,
\begin{equation}
    I_{NV-ESR} = \exp \bigg[\frac{-2\pi \mu_0 \mu_B^2 g_{NV} g_B T}{9\sqrt{3}\hbar}  n \:   \bigg\langle \sin^2 \frac{\theta}{2} \bigg \rangle_L \bigg]
    \label{eq:deerIntViktor}
\end{equation}
where $\mu_0$ is the vacuum permeability, $\mu_B$ is the Bohr magneton, $g_{NV}$ is the $g$-value of the NV center, $g_B$ is the g value of target spins, $T$ is the time for phase to accumulate after application of the DEER pulse, $\hbar$ is the reduced Planck constant, and $n$ is the concentration of target spins.
The $\langle \sin^2 \frac{\theta}{2} \rangle_L$ term represents the effective population inversion of the DEER pulse given as,~\cite{Salikhov1981}
\begin{eqnarray}
\bigg\langle \sin^2 \frac{\theta}{2} \bigg \rangle_L =  \nonumber
\int_{-\infty}^{+\infty}
\frac{\Omega^2}{(\xi -\omega)^2 +\Omega^2} \\
\times \sin^2\bigg(\sqrt{(\xi -\omega)^2 + \Omega^2 }\frac{t_p}{2}\bigg)
L(\xi; \Delta \omega)d\xi \label{eq:SinSqZeta}
\end{eqnarray}
where $\Omega$ is the Rabi frequency of the target spins, $\omega$ is the frequency of MW2, $t_p$ is the applied pulse length, and $L(\xi; \Delta \omega)$ is an intrinsic ESR line of P1 spins where $\Delta \omega$ represents the linewidth.
Therefore, Eq.~\ref{eq:SinSqZeta} includes the effects of the MW excitation and the ESR line on the NV-ESR signal.
The P1 Rabi data were simulated with Eq.~\ref{eq:deerIntViktor} by fixing $T$ while allowing $n$ to vary.
As shown in Fig.~\ref{fig:DEER}(b), the simulations were found to be in good agreement with the experiments.
We found that NV-ESR intensity depends on the value of $T$.
As shown in Fig.~\ref{fig:DEER}(b), NV-ESR with $T=2.0$ $\mu$s exhibits a high intensity contrast between $\pi$ and $\pi/2$ pulses.

\begin{figure}
\includegraphics[scale=0.9]{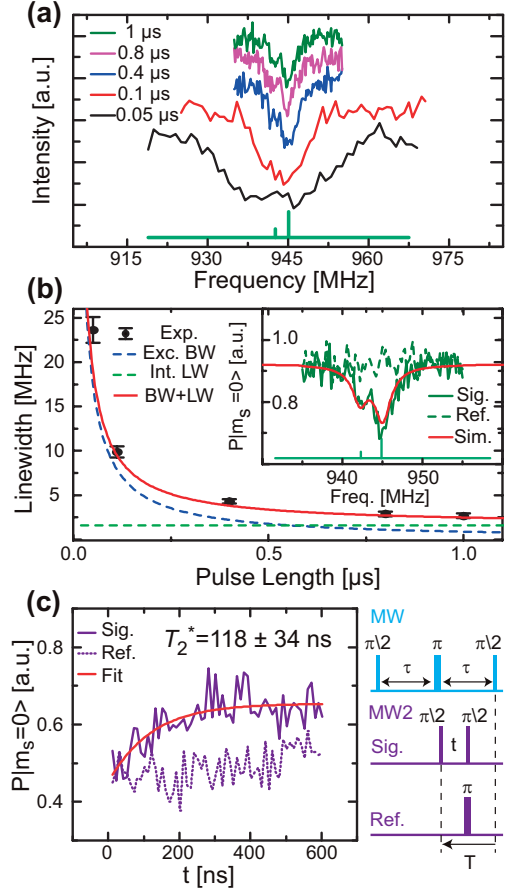}
\caption{Dependence of DEER pulse length on NV-ESR linewidth.
(a) NV-ESR spectra taken using various $\pi$-pulse lengths; $\pi$-pulse lengths are indicated in the legend.
(b) The NV-ESR linewidth as a function of the pulse length.
The red solid line is the result of a nonlinear least squares regression using Eq.~\ref{eq:deerIntViktor} and $\Delta \omega = 1.6$ (90 $\%$ confidence bounds of $(1.4, 1.8)$) MHz. Fitting was done with weights $1/\sigma^2$.
The blue and green dashed lines show partial contributions.
The blue dashed line is the MW excitation bandwidth, while the green dashed line shows $\Delta \omega$.
The inset graph shows the spectrum taken using a $\pi$-pulse of 1 $\mu$s.
The spectrum was normalized for the probability of the NV $\ket{0}$ state.
The simulated spectrum based on a linewidth of $1.6$ MHz is shown in red.
(c) Ramsey measurement using NV-ESR to measure $T_2^*$.
Pulses were applied 2 $\mu$s before the end of the sequence ($T=2$ $\mu$s).
The pulse sequence used for NV-ESR Ramsey is shown to the right.
The spacing between the pulses ($t$) was varied in the NV-ESR Ramsey measurement (Sig.).
A reference experiment was performed concurrently with the position of a single $\pi$-pulse being varied by $t$.
NV center (P1 center) pulse times were 40 (64) and 24 (32) ns, for the $\pi$ and $\pi/2$ pulses respectively.
The fit is shown in red.}
\label{Fig:DeerSpec}
\end{figure}
We seek to determine the origin of the observed linewidth in Fig.~\ref{fig:DEER}(a) by extracting the contribution from the MW excitation bandwidth.
The contribution is studied by analyzing the NV-ESR linewidth as a function of the DEER pulse length ($t_p$).
In the experiment, the MW power is adjusted to maintain a $\pi$-pulse for all pulse lengths.
As shown in Fig.~\ref{Fig:DeerSpec}(a), the spectrum narrows and the shape of the spectrum changes as the pulse length of the DEER pulse increases.
In order to characterize the linewidth, we fit each spectrum to a sum of two Lorentzians with resonance positions for all orientations of P1 centers ({\it i.e.}, $m_I=0$ transitions for the [111] orientation and other three orientations at $943$ and $945$ MHz, respectively).
The extracted full widths at half maximum (FWHM) are summarized in Fig.~\ref{Fig:DeerSpec}(b) where the linewidths strongly depend on $t_p$ below a $t_p$ of 0.4 $\mu$s.
To explain the dependence of the pulse length on the linewidth, we analyze contributions to the linewidth by fitting FWHM calculated using Eq.~\ref{eq:deerIntViktor} with the experimental FWHM where $\Delta \omega$ (Lorentzian linewidth) is a fit parameter.
As shown in Fig.~\ref{Fig:DeerSpec}(b), we found excellent agreement with the observed linewidths with $\Delta \omega = 1.6$ MHz with 90$\%$ confidence bounds of $(1.4,1.8)$ MHz.
In the figure, we also show partial contributions of the MW excitation and $\Delta \omega$ where the contribution of the MW excitation bandwidth is obtained by numerical calculation of FWHM using $L(\xi;\Delta \omega) = \delta(\xi)$ in Eq.~\ref{eq:deerIntViktor}.
This analysis verified that the MW excitation bandwidth is a major contribution of the NV-ESR linewidth when t$_p$ is shorter than $\sim$0.4 $\mu$s.
As shown in the inset of Fig.~\ref{Fig:DeerSpec}(b), the observed NV-ESR spectrum is well-explained by the simulation using Eq.~\ref{eq:deerIntViktor} with $\Delta \omega = 1.6$ MHz.
Moreover, the obtained high spectral resolution NV-ESR spectrum allows clear identification of the $m_I = 0$ P1 ESR signals for the [111] orientation and the non-[111] orientations which are separated by only 2 MHz.
This small splitting was not resolved in a previous experiment performed at a similar magnetic field.~\cite{Abeywardana2016}
The splitting is due to the contribution of the anisotropic hyperfine interaction comparable to the Zeeman energy at the low magnetic field which sets the ESR frequency from the non-[111] orientations to $945$ MHz at 33.4 mT while the [111] orientation remains at $943$ MHz.
We next confirm the nature of the intrinsic linewidth $\Delta \omega = 1.6$ MHz by comparing with the spin dephasing time ($T_2^*$).
$T_2^*$ relaxation time originates from an inhomogeneous distribution of ESR frequencies and represents the linewidth in many conventional ESR experiments.
Given the $\pi$ and $\pi/2$ pulse lengths with $T = 2$ $\mu$s as shown in Fig.~\ref{fig:DEER}(b), we perform a DEER Ramsey experiment to measure $T_2^*$.
To confirm the observed signal, a concurrent experiment varying the position of the $\pi$-pulse (Ref.) was performed with the sequence shown in Fig.~\ref{Fig:DeerSpec}(c).
We observed exponential behavior from the Ramsey measurement, as shown in Fig.~\ref{Fig:DeerSpec}(c).
The observed signal was then analyzed by fitting the data with Eq.~\ref{eq:deerIntViktor} where $\langle \sin^2 \frac{\theta}{2} \rangle_L = - \exp(t/T_2^*)$.
We observed a $T_2^*$ =  118 $\pm$ 34 ns from the analysis for NV1.
The value of $T_2^*$ corresponds to a FWHM of $2.7 \pm 1.0$ MHz, a value in reasonable agreement with the $\Delta \omega$ extracted from frequency measurements.
Furthermore, from the analysis of the NV-ESR intensity at 943 MHz, the detected magnetic dipole field ($B_{Dip}$) is $\approx$420 nT.~\cite{Abeywardana2016}
This strength of the magnetic field corresponds to an axially aligned single spin at a distance of $\sim$16 nm.

\begin{figure*}
    \centering
    \includegraphics{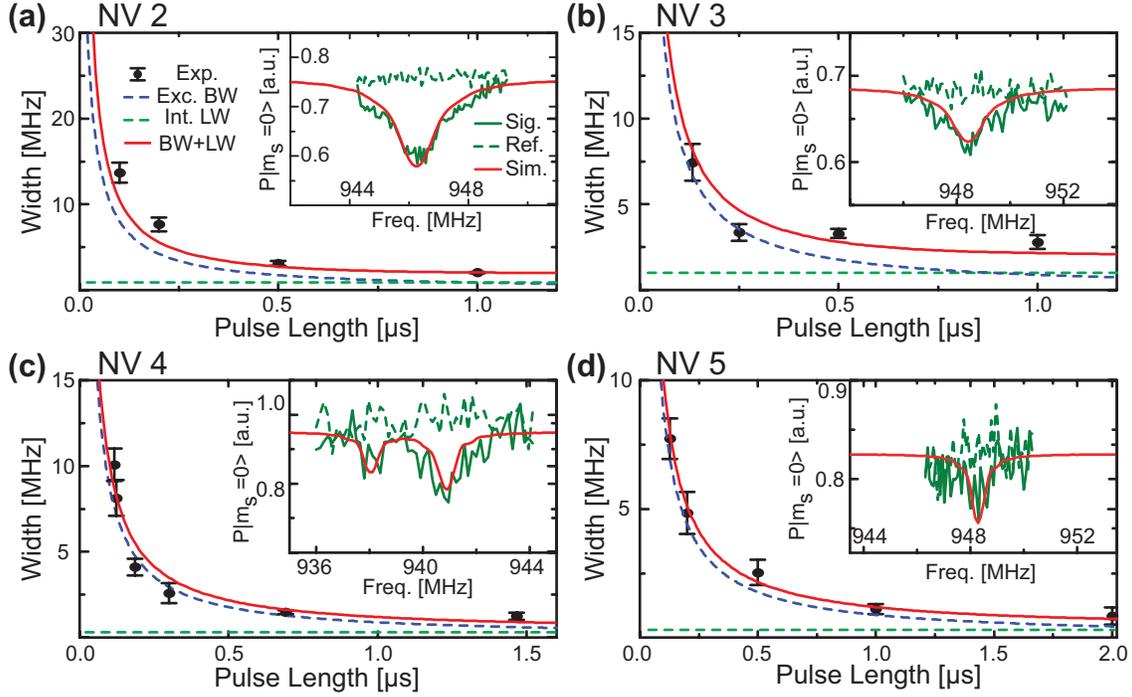}
    \caption{Dependence of DEER pulse length on NV-ESR linewidth for NVs 2-5. $\omega_0$ was set to the resonance position(s) for linewidth extraction as discussed in the main text.
    (a) Result of NV2.
    $\tau = 19.5$ $\mu$s and $B_0 = 37.7$ mT.
    The red solid line shows the fit result and the blue and green dashed lines show partial contributions from the MW excitation and $\Delta \omega$, respectively.
    From the fit, $\Delta \omega = 0.9 (0.9, 1.0)$ MHz was obtained.
    The inset graph shows the spectrum taken using a $\pi$-pulse of 1 $\mu$s (green) with the fitted spectrum shown in red.
    (b) Result of NV3.
    $\tau = 15$ $\mu$s and $B_0 = 37.8$ mT.
    From the fit, $\Delta \omega = 1.0 (0.5, 1.5)$ MHz was obtained.
    The inset graph shows the spectrum taken using a $\pi$-pulse of 1 $\mu$s.
    (c) Result of NV4.
    $\tau = 8.81$ $\mu$s and $B_0 = 32.7$ mT.
    Linewidth data was extracted using a sum of two equal-width Lorentzians with $\omega_1$ and $\omega_2$ set to be 941 and 943 MHz respectively.
    From the fit, $\Delta \omega = 0.3 (0.0, 0.6)$ MHz was obtained.
    The inset graph shows the spectrum taken using a $\pi$-pulse of 1.6 $\mu$s with the simulated spectrum shown in red.
    (d) Result of NV5.
    $\tau = 15$ $\mu$s and $B_0 = 37.8$ mT.
    From the fit, $\Delta \omega = 0.3 (0.2, 0.4)$ MHz was extracted.
    The inset graph shows the spectrum taken using a $\pi$-pulse of 2 $\mu$s.}
    \label{fig:SINV}
\end{figure*}
Moreover, we investigate NV-ESR spectroscopy with other single NV centers (NV2-5).
As summarized in Fig.~\ref{fig:SINV}, NV2-5 also exhibit a strong pulse length dependence similar to that observed for NV1.
For NV2, $\Delta \omega$ of 0.9 (0.9, 1.0) MHz was observed in agreement with the measured $T_2^*$ of $240 \pm 125$ ns, as shown in Fig.~\ref{fig:SINV}(a).
The linewidth observed for NV3 ($\Delta \omega = 1.0$ MHz) was similar to that of NV2 (Fig.~\ref{fig:SINV}(b)).
The linewidth observed for NV4 and NV5 ($\Delta \omega = 0.3$ MHz) was similar in magnitude, but significantly smaller than NV1-3 (Fig.~\ref{fig:SINV}(c) and (d)).
For NV4, we measure the $m_I=0$ ESR transition and resolve a very narrow linewidth that allowed for clear resolution of the two peaks originated from the [111] and other orientations of P1 spins.
In NV5, the NV-ESR linewidth with a pulse length of 2 $\mu$s was only 0.3 (0.2, 0.4) MHz.
Overall, the results from NV1-5 provide clear examples of nanoscale ESR investigation of the inhomogeneity in ESR signals. This is shown by variation in ESR linewidths as measured by different NVs located within the same diamond crystal.
Moreover, the observed linewidths of NV-ESR are much smaller than that in HF ESR (see Fig.~\ref{fig:HF_ESR}).
This is most likely because of the significant difference in the sample size between the two experiments.
Conventional ESR obtains signal from all spins within the millimeter scale sample while the sample volume in NV-ESR is confined within several to a-few-tens of nanometers from the NV center.
This significantly smaller size of the sample volume limits the number of detected P1 spins.
In the present case, P1 ESR in the nanometer-scaled sample volume has significantly smaller inhomogeneity compared with conventional ESR.
As shown by the previous conventional ESR investigations, there are two major contributions to the P1 ESR linewidth; hyperfine couplings to $^{13}$C nuclear spin baths and magnetic dipole couplings to P1 spin baths.~\cite{Stepanov2016,vanWyk97}
When the P1 concentration is low, the ESR linewidth as narrow as $\sim$0.3 MHz is broadened by the hyperfine couplings to the $^{13}$C nuclear spin baths.
On the other hand, when the P1 concentration is high, the linewidth is broader due to the coupling of the P1 spin baths and depends on the P1 concentration.
Therefore the present result strongly suggests that the variation of observed P1 linewidths is due to inhomogeneity of densities and spatial configurations of P1 spin baths within the detected nanoscale volume.
Furthermore, for NV 4 and 5, the contribution from the P1 spin baths is negligible on the linewidths ($\sim$0.3 MHz) while the hyperfine coupling to $^{13}$C spin baths is the major contribution.

\section{Conclusion}
Within this article we investigated the nature of the NV-ESR linewidth by studying P1 ESR.
We found that the spectral resolution depends strongly on the length of the DEER pulse.
This was particularly evident when pulse lengths are shorter than 0.4 $\mu$s.
Upon using long pulse lengths, the minimum resolved linewidth was found to be limited by inhomogeneous broadening of P1 ESR ($T_2^*$-limit).
This linewidth was found to vary between NV centers, indicating spatial inhomogeneity of local magnetic fields surrounding each NV center.
Since NV-ESR is useful for investigation of many spin systems with single spin sensitivity, the ability to perform high-resolution NV-ESR is critical.
The present work provides important context into the improvement of NV-ESR spectral resolution.
In particular, we demonstrated resolution of a small ESR splitting (2 MHz) by improving the spectral resolution and identified dominant coupling between P1 and surrounding electron and nuclear spins.
Furthermore, the present technique will be applicable for various NV-ESR investigations including identification of multiple types of spins and study of spin dynamics.

 \section{Acknowledgements}
This work was supported by the National Science Foundation (DMR-1508661 and CHE-1611134), the USC Anton B. Burg Foundation and the Searle scholars program (ST).


\providecommand{\latin}[1]{#1}
\makeatletter
\providecommand{\doi}
  {\begingroup\let\do\@makeother\dospecials
  \catcode`\{=1 \catcode`\}=2 \doi@aux}
\providecommand{\doi@aux}[1]{\endgroup\texttt{#1}}
\makeatother
\providecommand*\mcitethebibliography{\thebibliography}
\csname @ifundefined\endcsname{endmcitethebibliography}
  {\let\endmcitethebibliography\endthebibliography}{}

\end{document}